# Concatenated Quantum Codes

E. Knill, R. Laflamme

knill@lanl.gov, Mail Stop B265
laflamme@lanl.gov, Mail Stop B288
Los Alamos National Laboratory
Los Alamos, NM 87545
and
Institute for Theoretical Physics
Univversity of California, Santa-Barbara, CA
933106-4030.

August 8, 1996

**Abstract**

One of the main problems for the future of practical quantum computing is to stabilize the computation against unwanted interactions with the environment and imperfections in the applied operations. Existing proposals for quantum memories and quantum channels require gates with asymptotically zero error to store or transmit an input quantum state for arbitrarily long times or distances with fixed error. In this report a method is given which has the property that to store or transmit a qubit with maximum error $\epsilon$ requires gates with error at most $c\epsilon$ and storage or channel elements with error at most $\epsilon$, independent of how long we wish to store the state or how far we wish to transmit it. The method relies on using concatenated quantum codes with hierarchically implemented recovery operations. The overhead of the method is polynomial in the time of storage or the distance of the transmission. Rigorous and heuristic lower bounds for the constant $c$ are given.

## 1  Introduction

Practical quantum computing and communication (QCC) requires protecting the desired states from unwanted interactions with the environment and errors in the applied operations. This requirement already exists in classical computing and communication and is solved by the use of error-correcting codes for memory and channels and by exploiting (explicitly or implicitly) very reliable majority logic for fault tolerant operations. Fullfilling this requirement for QCC appears to be substantially more difficult, but no longer impossible. There now are methods for error-correcting quantum memories and channels [1, 2, 3], a general technique for fault tolerant quantum computing [4], and a practical method for correcting for dominant operational errors in one proposed device [5]. It is now conceivable that a combination of device dependent methods and general error-correction techniques will lead to practical applications of QCC.

A common feature of the currently understood error-correction methods is that to achieve a given error in the output state requires arbitrarily low error in the applied operations, depending on the number of time steps and operations required to accomplish the desired transformation. The best result to date is Shor's method [4] requiring polylogarithmically small error. Here we demonstrate a method based on concatenated coding for storing or transmitting a qubit with error $\epsilon$ which only requires that storage or channel elements have error amplitude at most $\epsilon$ and operational error amplitudes are bounded by $c\epsilon$ for some constant $c$ *independent* of the number $n$ of time steps involved. This result holds for all $\epsilon \leq 1/120$ with $c \geq 1/180$. The method requires $O(n^\delta)$ many additional qubits per qubit transmitted, with $\delta$ dependent on the actual operational accuracy. A consequence of our concatenated coding method is that if it is possible to implement operations with maximum error bounded by a constant (to be determined), then the apparent time and distance limitations of quantum communication protocols based on independently transmitted qubits can be overcome. In principle the method can be implemented by simple quantum repeaters spaced at regular intervals in a quantum channel with sufficiently many parallel paths.

The paper is organized as follows: In Section 2 the basic concepts required for understanding concatenated quantum codes are reviewed. This includes the fundamentals of quantum coding, a formalism for discussing operational errors and their propagation based on superoperators and sufficient assumptions for proving the main property of concatenated quantum codes. In Section 2 the concatenated coding procedure is defined. The analysis of the procedure is given in Section 4.

## 2 Preliminaries

### 2.1 Quantum Codes

Our treatment of quantum codes is based on [6]. The basic system of interest is the qubit $\mathcal{Q}$, which is a two dimensional complex Hilbert space spanned by the *classical* (orthonormal) states $|0\rangle$ and $|1\rangle$[1]. The system consisting of $n$ qubits is denoted by $\mathcal{Q}^{\otimes n}$, the $n$-fold tensor product of $\mathcal{Q}$. Its classical basis consists of states of the form $|b\rangle$ with $b$ an $n$-bit binary string.

A quantum code for $\mathcal{Q}$ of length $n$ is a two-dimensional subspace $\mathcal{C}$ of $\mathcal{Q}^{\otimes n}$. The preferred basis of $\mathcal{C}$ is denoted by $|0_L\rangle$ and $|1_L\rangle$. For the purposes of error-correction, an abstract decomposition $\mathcal{Q}^{\otimes n} \simeq \mathcal{C} \otimes \mathcal{S} \oplus \mathcal{R}$ is given. Let the *syndrome* space $\mathcal{S}$ be spanned by $|i_S\rangle$. The decomposition is instantiated by the unitary map $\rho : \mathcal{C} \otimes \mathcal{S} \to \mathcal{Q}^{\otimes n}$, where we assume that $\rho(\mathcal{C} \otimes |0_S\rangle) = \mathcal{C}$. $\mathcal{C}$ and $\rho$ are $e$-error correcting if for every operator $U$ of the form $\otimes_{i=1}^{n} U_i$ with at most $e$ of the $U_i$ different from the identity,

$$U\rho(|i_L\rangle|0_S\rangle) = \rho(|i_L\rangle|\psi\rangle),$$

for some state $|\psi\rangle$ in the syndrome space. See [6] for representation independent characterizations of error-correcting quantum codes.

There are three types of operations that involve quantum codes: Encoding, recovery and decoding. All of these operations may involve ancilla qubits. The encoding operation $E$ unitarily transfers a state of a qubit to $\mathcal{C}$. The recovery operation $R$ is defined by $R(\rho(|i_L\rangle|\psi\rangle)) = \rho(|i_L\rangle|0_S\rangle)$. The recovery operation is not unitary on $\mathcal{Q}^{\otimes n}$, but can be extended to a unitary operation by using ancilla qubits in a fixed initial state to which the syndrome information can be transferred. For efficiency, measurement operations (in the classical basis) can be used on the ancilla qubits. The decoding operation $D$ can be described as a recovery operation followed by a unitary map which transfers the state of $\mathcal{C}$ to a qubit. Each of these operations is to be implemented using primitive one and two qubit operations subject to operational errors.

For concatenated quantum coding it suffices to have a short one-error correcting code with efficient implementations of the three operations on the codes. An example of a length five one-error correcting code is given in [9, 10].

---

[1] Generalizing this work to larger dimensional basic systems is straightforward.

## 2.2 Superoperators and Error Propagation

Errors in applying an intended unitary operation $U$ to a system $\mathcal{H}$ involve over-rotation as well as entanglement with the environment (e.g. decoherence and relaxation). Instead of basing our discussion on superoperators [7, 6] or error operators [8], we use a more direct approach. The actual unitary operation $U_a$ acts on $\mathcal{E} \otimes \mathcal{H}$ for some environment $\mathcal{E}$, with $\mathcal{E}$ in a pure initial state $|0_E\rangle$. Thus we can write

$$U_a|0_E\rangle|\psi\rangle = \sum_i \mathbf{a}_i A_i |\psi\rangle\,,$$

where the $\mathbf{a}_i$ are non-normalized states of $\mathcal{E}$ and the $A_i$ are linear operators. A superoperator representation of the effect of $U_a$ on $\mathcal{H}$ is obtained by requiring that the $\mathbf{a}_i$ are orthonormal. An error operator representation is obtained by requiring that the $A_i$ are members of a suitable orthogonal operator basis.

We define a *generalized operator* to be any sum of the form $\mathcal{A} = \sum_i \mathbf{a}_i A_i$. $\mathcal{A}$ is *generalized unitary* if it is derived from a unitary operator acting on $|0_E\rangle \otimes \mathcal{H}$. Applying $\mathcal{A}$ to a state $|\psi\rangle$ yields $\mathcal{A}|\psi\rangle = \sum_i \mathbf{a}_i A_i |\psi\rangle$. The environment is explicitly represented in this expression. This is primarily to allow manipulating linear expressions in $|psi\rangle$ and $\mathcal{A}$. The basis of the environment is irrelevant. This is usually made explicitly by considering density matrices rather than ensembles of states. The effect of $\mathcal{A}$ on a density matrix $\rho$ is given by

$$\mathcal{A}(\rho) = \sum_{i,j} \mathbf{a}_i^\dagger \mathbf{a}_j A_j \rho A_i^\dagger\,.$$

Two operators are considered equivalent if they have the same effect on density matrices. In particular, if $V$ is unitary on $\mathcal{H}$, then $\sum_i \mathbf{a}_i V$ is equivalent to a scalar multiple of $V$. In general $\sum_i \mathbf{a}_i A_i$ is equivalent to $\sum_i \mathbf{b}_i B_i$ iff there is a unitary map $U$ such that $\sum_i (U\mathbf{a}_i)_k A_i = \sum_i (\mathbf{b}_i)_k B_i$ for each $k$. The environment spaces may need to be extended by additional dimensions.) Here, the subscript $k$ refers ot the $k$'th component of the subscripted expression, and the sum is interpreted as a sum of matrices.

The *strength* $|\mathcal{A}|$ of $\mathcal{A}$ is the maximum length of $\mathcal{A}|\psi\rangle$. An explicit expression for the length is given by

$$|\mathcal{A}|\psi\rangle|^2 = \sum_{i,j} \mathbf{a}_i^\dagger \mathbf{a}_j \langle\psi|A_i^\dagger A_j|\psi\rangle\,.$$

$\mathcal{A}$ is derived from a unitary operator iff $|\mathcal{A}|\psi\rangle| = 1$ for each $|\psi\rangle$. Operators can be compared on the basis of fidelity. For our analysis of concatenated quantum coding it is more convenient to use another definition of error. Let $\mathcal{A}$ and $\mathcal{B}$ two generalized unitary operators. The error amplitude $E(\mathcal{A}, \mathcal{B})$ of $\mathcal{A}$ compared to $\mathcal{B}$ is the smallest $\epsilon$ such that we can write $\mathcal{A} = \lambda \mathcal{B}' + \mathcal{E}$, with $\mathcal{B}'$ equivalent to $\mathcal{B}$ and $|\mathcal{E}| \leq \epsilon$. For dimension two, the squared error is related by a constant to the various notions of errors based on fidelity [7, 6].

The need for considering error amplitudes rather than probabilities arises from the possibility of errors adding coherently. This implies that to exploit additive error propagation bounds requires using amplitudes. This yields correct worst case estimates. In many practical situations, errors add nearly classically and in fact, many algorithms are designed to avoid interference between errors. Thus it is not unreasonable to use the *dissipated error heuristic*, according to which we can consider error probabilities and use essentially classical reasoning to analyze the different error possibilities. However, it is important to realize that this is a heuristic which is strictly true only in special circumstances.

To discuss errors of operators on codes, we need to be able to compare the restrictions of operators to subspaces. Let $\mathcal{C}$ be a subspace of $\mathcal{H}$. The restriction of $\mathcal{A}$ to $\mathcal{C}$ is denoted by $\mathcal{A} \restriction \mathcal{C}$. The restriction's range may not agree with the domain and is usually larger. However, the notions of strength and error amplitude still apply.

In the remainder of this section we state the properties of error amplitudes and propagation required for the formal analysis of concatenated quantum coding.

In the definition of error amplitude, we can assume that $\lambda \leq 1$.

**Lemma 2.1.** *Let $\mathcal{A}$ and $\mathcal{B}$ be generalized unitary operators with identical domains. Suppose that $\mathcal{A} = \lambda \mathcal{B} + \mathcal{E}$ where $|\mathcal{E}| \leq \epsilon$. Then $\mathcal{A} = \lambda' \mathcal{B} + \mathcal{E}'$ with $|\lambda'| \leq 1$ and $|\mathcal{E}'| \leq \epsilon$.*

*Proof.* Suppose that $|\lambda| > 1$ and $|\psi\rangle$ is in the common domain of $\mathcal{A}$ and $\mathcal{B}$. Let $\lambda' = \frac{\lambda}{|\lambda|}$) and $\mathcal{E}' = \mathcal{E} + \lambda(1 - \frac{1}{|\lambda|})\mathcal{B}$. Let $\psi_A = \mathcal{A}|\psi\rangle$, $\psi_B = \frac{\lambda}{|\lambda|}\mathcal{B}|\psi\rangle$, $\Delta\psi_B = \lambda(1 - \frac{1}{|\lambda|})\psi_B$, $\psi_E = \mathcal{E}|\psi\rangle$ and $\psi_{E'} = \mathcal{E}'|\psi\rangle$. (States without the surrounding $|\rangle$ notation are potentially not normalized.) Because $\mathcal{A}$ and $\mathcal{B}$ are generalized unitary, $|\psi_A| = |\psi_B| = 1$. We have $\psi_{E'} = \psi_E + \Delta\psi_B = \psi_A - \psi_B$. It suffices to show that $|\psi_{E'}| \leq |\psi_E|$ to deduce that $|\mathcal{E}'| \leq \epsilon$. A simple geometric argument can be used.

$$\begin{aligned} |\psi_E|^2 &= |\psi_{E'} - \Delta\psi_B|^2 \\ &= |\psi_{E'}|^2 + |\Delta\psi_B|^2 - 2\operatorname{Re}\psi_{E'}^\dagger\Delta\psi_B\,. \\ -\psi_{E'}^\dagger\Delta\psi_B &= (\psi_B - \psi_A)^\dagger\Delta\psi_B \\ &= |\Delta\psi_B| - \psi_A^\dagger\Delta\psi_B\,. \end{aligned}$$

Since $|\psi_A| = 1$, the real part of the last expression is non-negative. Consequently $|\psi_E|^2 \geq |\psi_{E'}|^2$, as desired. $\square$

Let $\mathcal{A} = \sum_i \mathbf{a}_i$ and $\mathcal{B} = \sum_i \mathbf{b}_i$. Composition of generalized operators is defined by

$$\mathcal{AB} = \sum_{i,j} \mathbf{a}_i \otimes \mathbf{b}_j A_i B_j\,.$$

This assumes that the two environments associated with the operators are independent.

**Lemma 2.2.** *Let $\mathcal{A}_i, \mathcal{B}_i$, $i = 1,2$ be generalized unitary operators and $\mathcal{C}$ a subspace of $\mathcal{H}$. If $\mathcal{C}$ is an invariant subspace of $\mathcal{B}_1$ and $E(\mathcal{A}_i \upharpoonright \mathcal{C}, \mathcal{B}_i \upharpoonright \mathcal{C}) \leq \epsilon_i$, then*

$$E(\mathcal{A}_2\mathcal{A}_1 \upharpoonright \mathcal{C}, \mathcal{B}_1\mathcal{B}_2 \upharpoonright \mathcal{C}) \leq \epsilon_1 + \epsilon_2\,.$$

*Proof.* Write $\mathcal{A}_i \upharpoonright \mathcal{C} = \lambda_i \mathcal{B}'_i \upharpoonright \mathcal{C} + \mathcal{E}_i \upharpoonright \mathcal{C}$ with $|\mathcal{E}_i \upharpoonright \mathcal{C}| \leq \epsilon_i$ and $|\lambda_i| \leq 1$. By choosing $\mathcal{E}_i$ appropriately on the orthogonal complement of $\mathcal{C}$, we can assume that $\mathcal{A}_i = \lambda_i \mathcal{B}'_i + \mathcal{E}_i$. Using $\mathcal{B}_1 \mathcal{C} = \mathcal{C}$, we get

$$\begin{aligned} \mathcal{A}_2\mathcal{A}_1 \upharpoonright \mathcal{C} &= (\lambda_1\lambda_2 \mathcal{B}'_2\mathcal{B}'_1 + \mathcal{A}_2\mathcal{E}_1 + \lambda_1\mathcal{E}_2\mathcal{B}_1) \upharpoonright \mathcal{C} \\ &= \lambda_1\lambda_2\mathcal{B}'_2\mathcal{B}'_1 \upharpoonright \mathcal{C} + \mathcal{A}_2\mathcal{E}_1 \upharpoonright \mathcal{C} + \lambda_1\mathcal{E}_2\mathcal{B}'_1 \upharpoonright \mathcal{C} \\ &= \lambda_1\lambda_2\mathcal{B}'_2\mathcal{B}'_1 \upharpoonright \mathcal{C} + \mathcal{A}_2(\mathcal{E}_1 \upharpoonright \mathcal{C}) + \lambda_1(\mathcal{E}_2 \upharpoonright \mathcal{C})(\mathcal{B}'_1 \upharpoonright \mathcal{C})\,. \end{aligned}$$

$\mathcal{B}'_2\mathcal{B}'_1$ is equivalent to $\mathcal{B}_2\mathcal{B}_1$, $|\mathcal{A}_2(\mathcal{E}_1 \upharpoonright \mathcal{C})| \leq \epsilon_1$ (since $\mathcal{A}_2$ is generalized unitary), and $|\lambda_1(\mathcal{E}_2 \upharpoonright \mathcal{C})(\mathcal{B}_1 \upharpoonright \mathcal{C})| \leq \epsilon_2$ (since $\mathcal{B}_1$ is generalized unitary and $|\lambda_1| \leq 1$). $\square$

**Lemma 2.3.** *Let $\mathcal{A}$, $\mathcal{B}_1$, $\mathcal{B}_2$ and $\mathcal{R}$ be generalized unitary operators. Suppose that $E(\mathcal{B}_2, \mathcal{B}_1) \leq \epsilon$ and $E(\mathcal{B}_1\mathcal{A} \upharpoonright \mathcal{C}, \mathcal{R} \upharpoonright \mathcal{C}) \leq \delta$. Then $E(\mathcal{B}_2\mathcal{A} \upharpoonright \mathcal{C}, \mathcal{R} \upharpoonright \mathcal{C}) \leq \epsilon + \delta$.*

*Proof.* Write $\mathcal{B}_2 = \lambda_2 \mathcal{B}'_1 + \mathcal{E}_2$ with $|\mathcal{E}_2| \leq \epsilon$ and $\mathcal{B}_1\mathcal{A} = \lambda_1 \mathcal{R}' + \mathcal{E}_1$ with $|\mathcal{E}_1 \upharpoonright \mathcal{C}| \leq \delta$. Then

$$\begin{aligned} \mathcal{B}_2\mathcal{A} &= (\lambda_2\mathcal{B}'_1 + \mathcal{E}_2)\mathcal{A} \\ &= \lambda_2\mathcal{B}'_1\mathcal{A} + \mathcal{E}_2\mathcal{A}\,. \end{aligned}$$

$\mathcal{B}'_1\mathcal{A}$ is equivalent to $\mathcal{B}_1\mathcal{A}$. Using the characterization of equivalence, one can see that $\mathcal{B}'_1\mathcal{A} = \lambda_1\mathcal{R}'' + \mathcal{E}'_1$ with $\mathcal{R}''$ equivalent to $\mathcal{R}$ and $\mathcal{E}'_1$ equivalent to $\mathcal{E}_1$. Thus
$$\mathcal{B}_2\mathcal{A} = \lambda_1\lambda_2\mathcal{R}'' + \lambda_2\mathcal{E}'_1 + \mathcal{E}_2\mathcal{A}.$$
The result follows by bounding the strength of the last two summands. □

We will also make use of the fact that the error is decreasing under elimination of ancilla systems.

**Lemma 2.4.** *Let $\mathcal{A}$ and $\mathcal{B}$ be generalized unitary operators on $\mathcal{H}_1 \otimes \mathcal{H}_2$. Let $\mathcal{A}_1$ and $\mathcal{B}_1$ be the generalized operators induced on $\mathcal{H}_1$ by considering $\mathcal{H}_2$ as part of the environment and restricting the operators to a subspace of the form $\mathcal{H}_1 \otimes |0\rangle$. Then $E(\mathcal{A}, \mathcal{B}) \leq \epsilon$ implies $E(\mathcal{A}_1, \mathcal{B}_1) \leq \epsilon$.*

*Proof.* It suffices to observe that the strength of an operator is non-increasing under restriction. □

## 2.3 Assumptions

Without making assumptions on how errors occur it is not possible to prove nontrivial results on error-correction. To obtain the main result for concatenated quantum codes we make three assumptions. The first is embedded in the qubit formalism and requires that for all practical purposes, the physical system which implements a qubit has access to only the two dimensional Hilbert space described by the qubit. This is called the *no leakage* assumption. An example of a system which without modification does not usually satisfy this assumption is a photon, with $|0\rangle$ and $|1\rangle$ represented by horizontal and vertical polarizations, respectively. Photon's tend to be scattered or absorbed and thus lost to the computation. If the actual systems have more than two degrees of freedom and leakage does occur, this can in principle be fixed by returning the leaked amplitude to the qubit before each coding operation. This does not need to be done perfectly, provided the other two assumptions are satisfied. Consequently, the no leakage assumption is useful primarily for simplififying the analysis of errors of specific codes.

The second assumption is that in each time step, independent qubits evolve independently. This is called the *local independence* assumption. This means that in each time step, we can partition the qubits into disjoint sets $P_i$ of one or two qubits (according to the primitive operations we wish to apply in parallel), where the qubits in $P_i$ are operated on by a generalized unitary operator $\mathcal{A}_i$. The overall effect of the step is to apply $\otimes_i \mathcal{A}_i$.

The third assumption is that errors of sequential operations are independent. This is called the *sequential independence* assumption, and is implicit in using composition of the generalized operators of each time step to obtain the final state.

Weakenings of these assumptions are possible but complicate the analysis.

## 3 Concatenated Quantum Codes

Although quantum error-correcting codes can reduce the effect of local interactions such as decoherence, a one-time use of such a code cannot recover a state after an arbitrary amount of time. The problem is that most interactions which destroy the state are time-dependent with a typical time scale for total loss. The effective error amplitude introduced by the interaction can be approximately modelled by a function of the form $1 - e^{-t/\tau}$. If $t \gg \tau$, there his no hope of recovering the state by any single use of a quantum error-correcting code.

One method for extending the lifetime of a state is by applying recovery operations to the coded state sufficiently frequently. Suppose that an error free recovery operation is applied every $t$ time units and that the code is $e$-error-correcting. The error rate after $t$ time is $1 - e^{-t/\tau}$, which is reduced by recovery to at most $c(1 - e^{-t/\tau})^{e+1}$ for some constant $c$. Provided that the total time $T$ satisfies $\frac{T}{t}c(1 - e^{-t/\tau})^{e+1} \ll 1$, the state still has high fidelity after $T$ time. Clearly, to increase the survival time of the state, the interval $t$ has to be reduced or a code correcting more errors must be used. Furthermore, if the recovery operation is not error free, residual errors will accumulate and limit the total time for which the state can be maintained. See [4] for a method of minimizing, but not eliminating the residual errors.

Concatenated quantum coding provides a simple method for eliminating the requirement for arbitrarily small operational errors during recovery operations. They are a demonstration of the ability to chain many error-prone operations in such a way that the final error is not much larger than that of a single operation. The basic idea is to hierarchically code each qubit and interlace the procedure with recoveries in such a way that errors do not propagate as they would using simple repeated recovery operations.

Concatenated quantum coding depends on a hierarchical implementation of a fixed error-correcting code. Let $\mathcal{C} \subseteq \mathcal{Q}^{\otimes l}$ be a two dimensional $e$-error-correcting code of length $l \geq 2$ with encoding operation $\mathcal{E}$, recovery operation $\mathcal{R}$ and decoding operation $\mathcal{D}$. Assume without loss of generality that $\mathcal{E}, \mathcal{D}, \mathcal{R} : \mathcal{Q}^{\otimes l} \to \mathcal{Q}^{\otimes l}$ with $\mathcal{E}|i\rangle|0\rangle = |i_L\rangle$, and for every correctable operator $A$, $\mathcal{R}A|i_L\rangle \propto |i_L\rangle$ and $\mathcal{D}A|i_L\rangle = |i\rangle|0\rangle$. Note that $\mathcal{R}$ and $\mathcal{D}$ must be generalized unitary operators. Let $r$ be a *repetition factor*, $r \geq 2$. The repetition factor is taken as large as reasonable subject to constraints to be determined by the analysis. The length of the code is largely irrelevant (except for overhead considerations), what matters is how much error per qubit can be corrected with a good overall error after recovery.

We recursively define concatenated coding procedures $\text{CCP}_{r,h}$ for each level $h$. The lowest level procedure $\text{CCP}_{r,1}$ consists of simple iterated recoveries between an encoding and a decoding operation. That is, $\text{CCP}_{r,1}$ begins with one qubit, encodes it to $l$ qubits using $\mathcal{C}$, applies a recovery procedure to the code $r - 1$ times and finally decodes it back to a single qubit[2]. In between recovery operations, we can either just wait for a certain time interval, or transmit each qubit over some distance.

$\text{CPP}_{r,1}(\underline{q})$
**Input:** A qubit $\underline{q}$, in a state which may be entangled with other systems.
**Output:** The qubit $\underline{q}$ in a state close to the input state.

>   $\underline{a} \leftarrow |0\rangle \in \mathcal{Q}^{\otimes l-1}$
>     **C:** *The underline notation $\underline{x}$ is used to explicitly indicate that the register $x$ may be in a non-classical state and entangled with other systems.*
>   $\mathcal{E}(\underline{qa})$
>     **C:** *Apply the encoding operation to $\underline{q}$ and $\underline{a}$. The new state is in $\mathcal{C} \subseteq \mathcal{Q}^{\otimes l}$.*
>   **for** $i = 1$ **to** $r - 1$
>     Wait or transmit each qubit of $\underline{qa}$.
>     $\mathcal{R}(\underline{qa})$
>   Wait or transmit each qubit of $\underline{qa}$.
>   $\mathcal{D}(\underline{qa})$
>   **dissipate** $\underline{a}$
>     **C:** *The state of $\underline{a}$ can be measured or simply discarded. The qubits of $\underline{a}$ can be reset and used again if so desired. To satisfy the independence assumption, it is important that the (former) contents of $\underline{a}$ have no effect on the remainder of the computation.*

---

[2] The repetition factor is $r$ because the final decoding operation is a special form of the recovery operation, so in effect, $r$ recovery operations are used.

**return** $\underline{q}$

The higher level procedures $\text{CCP}_{r,h+1}$ are defined recursively, using a procedure like $\text{CCP}_{r,1}$, but with the next lower level applied to each qubit between recoveries. That is $\text{CCP}_{r,h+1}$ starts with one qubit, encodes it using the code, applies $\text{CCP}_{r,h}$ to each of the qubits of the code and recovers the code $r-1$ times, applies $\text{CCP}_{r,h}$ to each qubit again and finally decodes the state to one qubit.

$\text{CPP}_{r,h+1}(\underline{q})$
**Input:** A qubit $\underline{q}$, in a state which may be entangled with other systems.
**Output:** The qubit $\underline{q}$ in a state close to the input state.

    $\underline{a} \leftarrow |0\rangle \in \mathcal{Q}^{\otimes l-1}$
    $\mathcal{E}(\underline{qa})$
    $\underline{b} \equiv \underline{qa}$
        **C:** $\underline{b}$ *is defined to refer to the $l$ qubits consisting of $\underline{q}$ and $\underline{a}$.*
    **for** $i = 1$ **to** $r - 1$
       **foreach** $i \in \{1, \dots, l\}$
          $\text{CPP}_{r,h}(\underline{b_i})$
            **C:** $\underline{b_i}$ *refers to the $i$'th qubit of $\underline{b}$.*
       $\mathcal{R}(\underline{b})$
    **foreach** $i \in \{1, \dots, l\}$
       $\text{CPP}_{r,h}(\underline{b_i})$
    $\mathcal{D}(\underline{qa})$
    **dissipate** $\underline{a}$
    **return** $\underline{q}$

A possible modification of $\text{CCP}_{r,h}$ includes adding waiting periods before and/or after each call to the next lower level procedure. The maximum length of these waiting periods is determined by the errors introduced by these periods and the correction capability of the implementation of the code.

## 4 Analysis

### 4.1 Error Propagation

**Theorem 4.1.** *Let $\mathcal{E}'$, $\mathcal{R}'$ and $\mathcal{D}'$ be implementations of the encoding, recovery and decoding operations. Assume the following:*

(1) $E(\mathcal{E}', \mathcal{E}) \leq \epsilon_c$.

(2) *If $\mathcal{I}$ is a generalized unitary operator on one qubit with $E(\mathcal{I}, I) \leq \epsilon_d$, then $E(\mathcal{R}'\mathcal{I}^{\otimes l} \upharpoonright \mathcal{C}, I \upharpoonright \mathcal{C}) \leq \epsilon_c$ and $E(\mathcal{D}'\mathcal{I}^{\otimes l} \upharpoonright \mathcal{C}, \mathcal{D} \upharpoonright \mathcal{C}) \leq \epsilon_c$.*

(3) *The error operator acting during the waiting or transmission period in $\mathrm{CCP}_{r,h}$ is of the form $\mathcal{M} = \mathcal{I}^{\otimes l}$, with $E(\mathcal{I}, I) \leq \epsilon_d$. $\mathcal{M}$ is called the channel error.*

(4) $r + 1 \leq \epsilon_d/\epsilon_c$.

*Then the error amplitude of $\mathrm{CCP}_{r,h}$ compared to the identity is at most $(r+1)\epsilon_c$.*

The fact that codes with the requisite properties exist is discussed in Section 4.3. A more detailed error analysis for a specific code is given in Section 4.4.

*Proof.* The proof is by induction on the level of the procedure. By the error propagation properties, the errors in the encoding operation, each of the $r - 1$ composition of channel error and recovery operation, and the final channel error followed by decoding can be added to get the error in the overall operation on the input qubit. Thus the error amplitude of $\mathrm{CCP}_{r,1}$ is at most $(r+1)\epsilon_c \leq \epsilon_d$.

The procedure $\mathrm{CCP}_{r,h+1}$ behaves exactly like $\mathrm{CCP}_{r,h}$ except that the waiting time or channel transmission is replaced by an application of $\mathrm{CCP}_{r,h}$. We assume inductively that the error of $\mathrm{CCP}_{r,h}$ is at most $\epsilon_d$. By the locality assumption, it follows that the error between each of the recovery operations is of the right form to deduce that the overall error is again bounded by $(r+1)\epsilon_c \leq \epsilon_d$. □

### 4.2 Overhead Requirements

Suppose that we wish to use concatenated quantum coding for storing or transmitting a state for $n$ multiples of the time or distance for which a qubit is subject to an interaction with the environment of error amplitude $\epsilon_d$. The most important resource requirements are the maximum number of qubits that are stored or transmitted, the total number of encoding, recovery or decoding operations and the number of parallel encoding, recovery or decoding steps. Consider $\text{CCP}_{r,h}$. By induction on $h$, it can be seen that the number $n$ of waiting/transmission periods is given by $r^h$. This requires $c(h) = l^h$ many qubits (not including ancillas that may be required for some of the operations). The number of parallel operations $p(h)$ satisfies $p(1) = r+1$, and $p(h+1) = r+1+rp(h)$. Thus $p(h) = (r+1)(r^h-1)/(r-1) < r^h(r+1)/(r-1)$. The total number $e(h)$ of operations satisfies $e(1) = r+1$ and $e(h+1) = r+1+rle(h)$. Thus $e(h) = (r+1)((rl)^h - 1)/(rl-1) \leq (rl)^h$. By expressing these relationships in terms of $n$ the following result is obtained:

**Theorem 4.2.** *To implement $\text{CCP}_{r,h}$ with $n$ waiting/transmission periods using a code of length $l$ requires $n^{\log_r l}$ qubits, less than $\frac{r+1}{r-1}n$ parallel operations and less than $n^{1+\log_r(l)}$ basic encoding, recovery and decoding operations.*

### 4.3 Existence of Suitable Codes

Any $e$-error correcting quantum code can be used for the code underlying the CPP, provided the basic operations can be implemented accurately enough. The critical requirement that must be met is $r + 1 \leq \epsilon_d/\epsilon_c$. The smallerst $r$ of interest is 2, so that it is necessary to use codes where $\epsilon_d/\epsilon_c \geq 3$.

Let $\mathcal{C}$ be an $e$-error correcting code of length $l$, with $e \geq 1$. Assume first that the basic operations on the code are implemented perfectly. It can be shown that a local error of amplitude $\epsilon$ per qubit is reduced to $O(\epsilon^{e+1})$ by the code.

**Theorem 4.3.** *Let $\mathcal{I}$ be a generalized unitary operator with $E(\mathcal{I}, I) \leq \epsilon$. If $\mathcal{R}$ is the recovery or the decoding operator for $\mathcal{C}$, then*

$$\begin{aligned} E(\mathcal{RI}^{\otimes l} \upharpoonright \mathcal{C}, I \upharpoonright \mathcal{C}) &\leq \sum_{i=e+1}^{l} \binom{l}{i} \epsilon^i \\ &\leq 2 \binom{l}{e+1} \epsilon^{e+1} \\ &\qquad \text{if } \epsilon \leq \tfrac{e+2}{2(l-e-1)}. \end{aligned}$$

There are two important differences between this bound and the usual one obtained for classical error correction. The first is that we are concerned with error amplitudes rather than probabilities. The second is that in the classical bound for the error probability, the factor $\epsilon^i$ is replaced by $\epsilon^{2i}(1-\epsilon^2)^{l-i}$.

*Proof.* Write $\mathcal{I} = \mathbf{a}I + \mathcal{E}$ with $|\mathbf{a}| \leq 1$ and $|\mathcal{E}| \leq \epsilon$. For $U \subseteq \{1,\ldots,l\}$, let $\mathcal{I}_U = \otimes_{i=1}^{l} \mathcal{A}_i$ with $\mathcal{A}_i = \mathbf{a}I$ for $i \notin U$, and $\mathcal{A}_i = \mathcal{E}$ otherwise. Then $\mathcal{I}^{\otimes l} = \sum_U \mathcal{I}_U$. If $|U| \leq e$, then $\mathcal{RI}_U \restriction \mathcal{C} = \mathbf{r}_U I \restriction \mathcal{C}$. Thus

$$\begin{aligned}\mathcal{RI}^{\otimes l} \restriction \mathcal{C} &= \sum_U \mathcal{RI}_U \restriction \mathcal{C} \\ &= \sum_{|U| \leq e} \mathbf{r}_U I \restriction \mathcal{C} + \sum_{|U| \geq e} \mathcal{RI}_U \restriction \mathcal{C}\,.\end{aligned}$$

The strength of $\mathcal{I}_U$ is bounded by $\epsilon^{|U|}$, by the usual tensor product rules for the strength of operators. Since the strength is subadditive, it follows that

$$I(\mathcal{RI}^{\otimes l} \restriction \mathcal{C}, I \restriction \mathcal{C}) \leq \sum_{i \geq e+1} \binom{l}{i} \epsilon^i\,.$$

The second inequality in the statement of the theorem follows by bounding the sum by a geometric series. □

If the the local error per qubit is $\epsilon_d$, and the error amplitude in the implementation of the recovery or decoding operation is $\epsilon_r$ on $\mathcal{Q}^{\otimes l}$, then the total error after recovery or decoding is at most $\epsilon_c = \epsilon_r + O(\epsilon_d^{e+1})$, with the constants determined by the parameters of the code. Clearly for sufficiently small $\epsilon_r$ and $\epsilon_d$, $\epsilon_d/\epsilon_c \geq 3$.

A nice feature of concatenated coding is that any code implemented with sufficiently high fidelity can be used, it does not need to correct any one type of error perfectly, only with low final error amplitude.

## 4.4 Example: The Five Qubit Code

Here is an explicit analysis of the behavior of concatenated quantum coding if the one-error correcting five qubit code of [9] is used. The analysis is based rigorously on amplitude errors. An analysis using the dissipation heuristic is obtained by replacing all the error amplitudes by error probabilities.

The number of primitive operations required to implement the recovery operator of the five qubit code is at most 30 [11]. The primitive operations required are controlled nots, Hadamard transforms, sign flips and bit flips. This is also an upper bound on the operations for encoding and decoding. Suppose that the error amplitude for the implemented primitive operation is $\epsilon_p$. Then the error amplitude for the basic operations on the code is at most $30\epsilon_p$. If each qubit is subjected to local error of amplitude at most $\epsilon_d$ (the memory or channel error), then the error after recovery is bounded by

$$\epsilon_c = 30\epsilon_p + 20\epsilon_d^2,$$

provided that $\epsilon_d \leq 1/2$. Ignoring the overhead, we can achieve final error $3\epsilon_c$ over any distance provided that $\epsilon_c/\epsilon_d \leq 1/3$. If the final error to be achieved is $\epsilon$, this gives the following set of constraints to be satisfied by $\epsilon_p$ and $\epsilon_d$:

$$\epsilon_d \leq 1/2 \tag{1}$$
$$\epsilon \geq 90\epsilon_p + 60\epsilon_d^2 \tag{2}$$
$$\epsilon_d \geq 90\epsilon_p + 60\epsilon_d^2. \tag{3}$$

We determine the maximum $\epsilon_p$ for which these inequalities can be solved. Given $\epsilon_d$, the maximum $\epsilon_p$ is determined by

$$\epsilon_p(\epsilon_d) = \min((\epsilon - 60\epsilon_d^2)/90, \epsilon_d(1 - 60\epsilon_d)/90).$$

Since the first expression in the minimum is decreasing, the maximum is either at the maximum of $\epsilon_d(1 - 60\epsilon_d)$, given by $\epsilon_d = 1/120$, or at the solution of $\epsilon - 60\epsilon_d^2 = \epsilon_d(1 - 60\epsilon_d)$, given by $\epsilon_d = \epsilon$. The former holds if $\epsilon \geq 1/120$. Thus we obtain the following result:

**Theorem 4.4.** *A qubit can be stored for arbitrary amounts of time or transmitted over arbitrary distances with polynomial overhead in time or distance and with a final error amplitude of $\epsilon$ provided that one of the following holds:*

(1) $\epsilon \geq 1/120$, *the basic storage or channel element has error amplitude at most $1/120$ and the primitive one and two qubit operations can be implemented with error amplitude at most $1/21600$.*

(2) $\epsilon < 1/120$, *the basic storage or channel element has error at most $\epsilon$ and the primitive one and two qubit operations can be implemented with error amplitude at most $\epsilon(1 - 60\epsilon^2)/90$.*

## 5 Conclusion

We have shown that under local and sequential independence assumptions, there is a threshold gate error which suffices for storing or transmitting a qubit for arbitrary times/distances at an overall error no larger than the error of a single memory or channel element and with polynomial overhead.

The minimum error amplitude requirement for success of the method translates to an error probability of about $.25 \; 10^{-8}$, which is out of reach of any forseeable technology. However, the dissipation heuristic gives a more optimistic estimate of $.5 \; 10^{-4}$, which seems more accessible. It should be possible to improve the error estimate for concatenated quantum coding by performing a more detailed analysis. An improvement may be obtainable by a more careful analysis of errors in the recovery operator, maybe exploiting the fault tolerant methods of [4, 11]. Another approach is to explicitly exploit knowledge of the physics of the implementation device to reduce error in operations. An example of this approach is [5].

Although the overhead of $n^c$ is not completely impractical, reduction of the constant $c$ in the exponent imposes more stringent accuracy requirements on the gates. Future work will be directed at reducing the overhead, ideally to a function polylogarithmic in $n$ with reasonable constants and exponents.

# 6 Acknowledgements

This work was performed under the auspices of the U.S. Department of Energy under Contract No. W-7405-ENG-36 and in part from the National Science Foundation under Grant No.PHY94-07194.